\documentclass[onecolumn,amsmath,amssymb,superscriptaddress,nofootinbib,11pt,a4paper]{revtex4}

\pdfoutput=1
\usepackage[T1]{fontenc}
\usepackage{CJK}
\usepackage{xcolor}
\usepackage{amsfonts}
\usepackage{epstopdf}
\usepackage{wrapfig} 
\usepackage{subfigure}
\usepackage{graphicx}  
\usepackage{dcolumn}   
\usepackage{bm}
\usepackage{float}
\usepackage[T1]{fontenc}
\usepackage{CJK}
\usepackage{xcolor}
\usepackage{amsfonts}
\usepackage{epstopdf}
\usepackage{wrapfig} 
\usepackage{subfigure}
\usepackage{graphicx}  
\usepackage{dcolumn}   
\usepackage{bm}
\usepackage{float}
\newcommand{\be}{\begin{equation}}
\newcommand{\ee}{\end{equation}}
 \newcommand{\bea}{\begin{eqnarray}}
 \newcommand{\ena}{\end{eqnarray}}

\newcommand{\bra}{\langle}
\newcommand{\ket}{\rangle}

\newcommand{\smalleq}{\leqslant}
\newcommand{\linear}{\propto}
\newcommand{\when}{\bigg\vert}

\newcommand{\notice}{\equiv}

\newcommand{\exd}{\mathrm{d}}

\newcommand{\pd}{\partial}

\newcommand{\diff}[2]{\frac{\exd{#1}}{\exd{#2}}}

\begin{document}
\title{Holographic Einstein rings of a black hole with a  global monopole}

\author{Xiao-Xiong Zeng}\email{xxzengphysics@163.com}
\affiliation{State Key Laboratory of Mountain Bridge and Tunnel Engineering, Chongqing Jiaotong University, Chongqing 400074, China}
\affiliation{Department of Mechanics, Chongqing Jiaotong University, Chongqing 400074, China}
\author{Li-Fang Li}{}\thanks{Corresponding author: lilifang@imech.ac.cn}
\affiliation{Center for Gravitational Wave Experiment, National Microgravity Laboratory, Institute of Mechanics, Chinese Academy of Sciences, Beijing 100190, China.}
\author{Peng Xu}
\affiliation{Center for Gravitational Wave Experiment, National Microgravity Laboratory, Institute of Mechanics, Chinese Academy of Sciences, Beijing 100190, China.}
\affiliation{Lanzhou Center of Theoretical Physics, Lanzhou University, No. 222 South Tianshui Road, Lanzhou 730000, China}

\begin{abstract}
The global monopole solutions which give rise to quite unusual physical phenomena have captured considerable attention. In this paper, we study the Einstein ring of the spherically symmetric AdS black hole solution with a global monopole based on the AdS/CFT correspondence. With the help of the given response function of the QFT on the boundary, we construct the holographic images of the black hole in the bulk. The Einstein rings on the images can be clearly observed. The holographic ring always appears with the concentric stripe surrounded when the observer located at the north pole, and an extremely bright ring when the observer is at the position of the photon sphere of the black hole. With the change of the observation position, this ring will change into a luminosity-deformed ring, or light points. Furthermore, we show that the monopole parameter has an effect on the brightness and the position of Einstein ring. All these results imply that the holographic images can be used as an effective tool to distinguish different types of black holes for fixed wave source and optical system.
\end{abstract}

\maketitle
 \newpage
\section{Introduction}
The AdS/CFT correspondence~\cite{Maldacena1998,Gubser1998,Witten1998}, also known as holographic duality, provides us an useful tool to research the gravity dual of quantum field theories(QFTs) and becomes very important in higer energy theoretical physics. One reason is that this correspondence can be used to understand strongly interacting field theory by mapping them to classical gravity~\cite{Aharony2000,DHoker0201253}. To be specific, the theory relates a gravity in a $(d+1)$-dimensional anti-de Sitter spacetime to a strongly coupled d-dimensional quantum field theory living on its boundary. Then the Ads/CFT correspondence is successfully extended to different domains, including the analysis of the strong coupling dynamics of QCD and the electroweak theories, the physics of black holes and quantum gravity, relativistic hydrodynamics or different applications in condensed matter physics~\cite{Hartnoll2009,McGreevy2010,Casalderrey1101,Kim1205,Adams1205,Gubser2008,Hartnoll2008,Hartnoll2008PRL,Herzog2009}. Therefore, the AdS/CFT correspondence provides strong support to study various physics topics.

Recently, the Event Horizon Telescope (EHT)~\cite{Falcke2000}, a global long baseline interferometry array observing at a wavelength of 1.3 mm, has captured the first image of supermassive black hole candidate in the center of the giant elliptical galaxy M87. The dark area inside the photo sphere, in which light rays have been absored by the black hole, is called black hole shadow~\cite{Akiyama2019}. The observed image is consistent with expectations for the shadow of a Kerr black hole as predicted by general relativity~\cite{Falcke2000}. Since then, the black hole shadows surrounded by different accretion models and their observational characteristics in the various gravity backgrounds have been also studied in~\cite{Narayan2019,Zeng2020,Qin2021,Saurabh2021,Zeng2022,He2022,Guo2021,Zeng2022Sci,Gralla2019,Li2021,Gan2021}. Considering these interesting phenomena, in this paper, we construct the holographic images of the black hole in the bulk from a given response function of the QFT on the boundary closely followed by~\cite{Hashimoto:2018okj, Hashimoto:2019jmw,Liu2022,Zeng:2023zlf}.

On the other hand, in recent years, the study of topological defects such as cosmic strings, domain walls, monopoles and textures have captured considerable attention and still remains one of the most active fields in modern physics. This is mainly due to their fascinating properties which give rise to a rich variety of quite unusual physical phenomena. As an intriguing one, a global monopole~\cite{Barriola1989} is simplest generated during the phase transition of a system composed of a self-coupling triplet scalar field whose original global $\mathrm{O}(3)$ symmetry is spontaneously broken to U(1). Since then, the physical properties of black hole incorporating global monopoles have been studied extensively~\cite{Jing1993,Yu1994,Dadhich1998,Li2002,
Gao2002,Jiang2006,Brihaye2006,Han2005,Wu2007,Chen2010,Lin2011}. However, the Einstein ring of a black hole with global monopoles in AdS spacetime still remains obscure. Motivated by this, in this paper, we focus our attention on investigating the holographic images of such model closely followed~\cite{Hashimoto:2018okj, Hashimoto:2019jmw,Liu2022,Zeng:2023zlf} with the AdS/CFT correspondence. The previous related studies have revealed the photon sphere induced Einstein ring structure does exist in holographic quantum matter, whereas the photon sphere varies according to the specific bulk dual geometry and the detailed behavior of the Einstein ring structure is also expected to vary. Here we investigate the behavior of the lensed response for a global monopole solution in the spherically symmetric AdS black hole and the possible effect of such monopole parameter on the resulting Einstein ring, which will further confirm that the appearance of such an Einstein ring structure may be used as a strong signal for the existence of its gravity dual.

Our paper is arranged as follows. In section~\ref{sec2}, we briefly review the thermodynamics of the global monopole solution in spherically symmeric AdS black hole. In section ~\ref{sec3}, we give a holographic setup of such model and analyze the lensed response function. With the optical system, we observe the Einstein ring in our model and compare our results with the optical approximation in section~\ref{sec4}. Our results shows that the position of photon ring obtained from the geometrical optics is full consistence with that of the holographic ring.
Section~\ref{sec5} is devoted to our conclusions.

\section{The global monopole solution in the spherically symmetric AdS black hole}
\label{sec2}
As stated in~\cite{Barriola1989}, monopoles formed as a result of a gauge-symmetry breaking are similar to elementary particles. Most of their energy is concentrated in a small region near the monopole core. In this paper, we would like to consider global monopoles, resulting from a global symmetry breaking. 
The asymptotically locally AdS black hole solution with a global monopole can be expressed as \cite{Barriola1989,Hu:2020lkg}
\begin{equation}
{ds}^2=-F\left({r}\right)d{t}^2+\frac{1}{F\left({r}\right)}d{r}^2+{r}^2\left(\text{d$\theta $}^2+\sin ^2\theta  \text{d$\phi $}^2\right),\label{metric1}
\end{equation}
where
\begin{equation}
F\left({r}\right)=1-b-\frac{2{m}}{{r}}+\frac{{r}^2}{l^2},\label{metric}
\end{equation}
in which, $b=8\pi  \eta ^2$ and $\eta $ is a parameter related to symmetry breaking, ${m}$ and ${q}$ correspond to mass and charge, $l$ is the radius of AdS space, which relates to the  cosmological parameters  with $\Lambda =-\frac{3}{l^2}$.

The laws of thermodynamics and the weak cosmic censorship conjecture of the above spherically symmetric AdS black hole solution with a global monopole have been carefully investigated in~\cite{Hu:2020lkg}. In the extended phase space, the first law of thermodynamics is valid, while the second law is violated for the extremal and near-extremal black holes. And the cosmic censorship conjecture is valid only for the extremal black hole, and is violated for the near-extremal black hole. In the following we will investigate the Einstein ring structure of such above model in the context of AdS/CFT.

\section{The holographic setup of Einstein image in the AdS black hole with a global monopole}
\label{sec3}
In the framework of AdS/CFT, we set up the holographic Einstein ring of AdS black hole with a global monopole. For simplicity, we set $z=r^{-1}$, then we have $F(r)=z^{-2}F(z)$. We reexpress the above metric~(\ref{metric1}) to be as
\begin{equation}
ds^2=\frac{1}{z^2}[-F(z)dt^2+\frac{dz^2}{F(z)}+d\Omega^2].
\end{equation}
The Hawking temperature in this case  is 
\begin{equation}
\label{temperature}
T=\frac{-b z_h^2+z_h^2+3}{4 \pi z_h}.
\end{equation}
For a massless particle in the scalar field, the corresponding dynamics is governed by the Klein-Gordon equation
\begin{equation}
\label{KG equation}
D_{a}D^{a}\Phi=0.
\end{equation}
To solve it by numerics in a more convenient manner, we perfer the ingoing Eddington coordinate, i.e.
\begin{eqnarray}
v\equiv t+z_{*}=t-\int \frac{1}{F(z)}dz.
\end{eqnarray}
As a result, the non-vanishing bulk background fields are transformed into the following smooth form
\begin{equation}
ds^2=\frac{1}{z^2}[-F(z)dv^2-2dzdv+d\theta^2+\sin^2\theta d\psi^2].
\end{equation}

With $\Phi=z \phi$, the asymptotic solution of equation~(\ref{KG equation}) near the AdS boundary $(z\rightarrow0)$ reads
\begin{equation}
\phi(v,z,\theta,\varphi)=J_\mathcal{O}(v,\theta,\varphi)
	+ \bra\mathcal{O}\ket z+O(z^2).
\end{equation}
By the holographic dictionary, $J_\mathcal{O}$ is interpreted as the source for the boundary field theory. And the corresponding expectation value of the dual operator $\bra\mathcal{O}\ket_{J_\mathcal{O}}$, namely the response function, is given by
\begin{align}\label{res1}	\bra\mathcal{O}\ket_{J_\mathcal{O}}=\bra\mathcal{O}\ket-\pd_v J_\mathcal{O},
\end{align}
where $\bra\mathcal{O}\ket$ corresponds obviously to the expectation value of the dual operator with the source turned off.

\begin{figure}[h]
	\centering
	\includegraphics[height=3.5in]{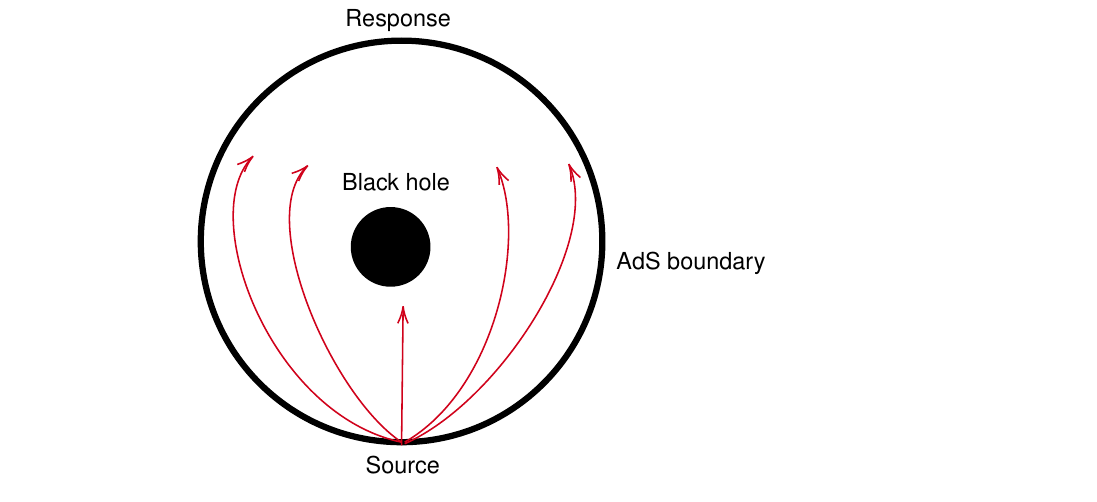}
	\caption{A monochromatic Gaussian source is located at a point on the AdS boundary, and its response is observed at another point on the same boundary. }\label{source1}
\end{figure}

Based on the AdS/CFT correspondence, we studied the holographic Einstein images of a global monopole solution in the bulk from a given response function on the side of the AdS boundary, where the response function is generated by the source on other side of the AdS boundary (see Fig.~\ref{source1}). We take an oscillatory Gaussian wave source $J_\mathcal{O}(v,\theta)$ on one side of the AdS boundary, and scalar waves generated by the source can propagate in the bulk. After the bulk of scalar wave reach the other side of AdS boundary, the corresponding response will be generated, i.e., the response function $|\bra \mathcal{O}\ket|$.

Here we employ the monochromatic and axisymmetric Gaussian wave packet centered on the south pole $\theta_0=\pi$ as the source
\begin{equation} \label{source}
	J_\mathcal{O}(v,\theta)=e^{-i\omega v}\frac{1}{2\pi \sigma^2}\exp\left[-\frac{(\pi-\theta)^2}{2\sigma^2}\right]=e^{-i\omega v}\sum_{l=0}^\infty {c}_{l0}Y_{l0}(\theta),
\end{equation}
where $\sigma$ is the width of the wave produced by the Gaussian source and $Y_{l0}$ is the spherical harmonics function.
We take the wave packet size $\sigma$ to be $\sigma\ll\pi$, and the coefficients of the spherical harmonics $Y_{l0}(\theta)$ can be calculated out as
\begin{equation}\label{coe}
 	c_{l0}=(-1)^l\sqrt{\frac{l+1/2}{2\pi}}\exp\left[-\frac12 (l+1/2)^2\sigma^2\right].
\end{equation}

With the source given by Eq. (\ref{source}), the corresponding bulk solution takes the following form
\begin{equation}\label{fieldDecomp}
	\phi(v,z,\theta)=e^{-i\omega v}\sum_{l=0}^\infty c_{l0}Z_l(z)Y_{l0}(\theta),
\end{equation}
where $Z_l$ satisfies the equation of motion
\begin{align}
z^2fZ_l''+z^2[f'+2i\omega]Z_l'+[-2f+zf'-z^2l(l+1)]Z_l=0,\label{zpart}
\end{align}
and its asymptotic behaviour near the AdS boundary goes like
\begin{equation}\label{expansion}
 	 Z_l=1+\bra\mathcal{O}\ket_l z+O(z^2).
\end{equation}
Similarly, the resulting response $\bra\mathcal{O}\ket_{J_\mathcal{O}}$ can be expressed as
\begin{equation}\label{resDecomp}
	\bra\mathcal{O}\ket_{J_\mathcal{O}}=e^{-i\omega v}\sum_{l=0}^\infty c_{l0}\bra\mathcal{O}\ket_{J_\mathcal{O} l} Y_{l0}(\theta)
\end{equation}
with
\begin{align}
	\bra\mathcal{O}\ket_{J_\mathcal{O} l}=\bra\mathcal{O}\ket_l+i\omega.
\end{align}

\begin{figure}[htbp]
	\centering
    \begin{minipage}{0.85\linewidth}
    \centering	\includegraphics[width=0.85\linewidth]{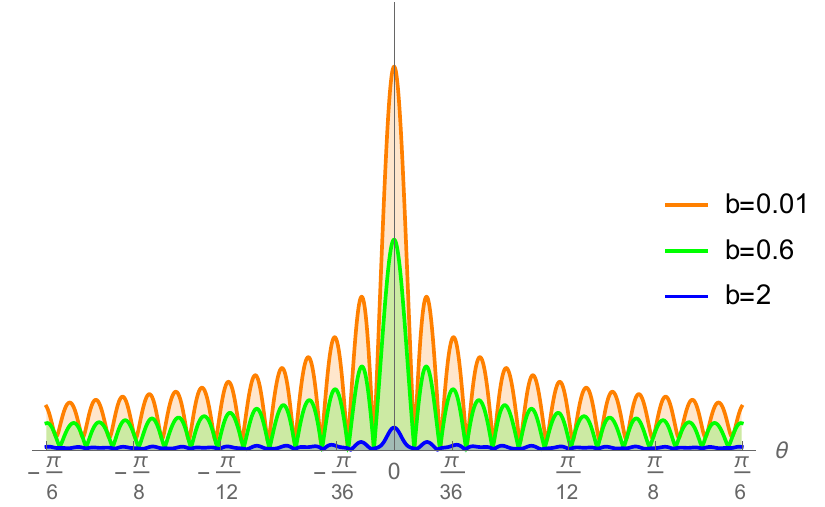}
	\caption{The amplitude of $|\bra \mathcal{O}\ket|$ for different $b$  with $z_h=1$ and $\omega=80$.}
    \label{amplitude1}
	\end{minipage}
 
    \begin{minipage}{0.85\linewidth}
	\centering
	\includegraphics[width=0.85\linewidth]{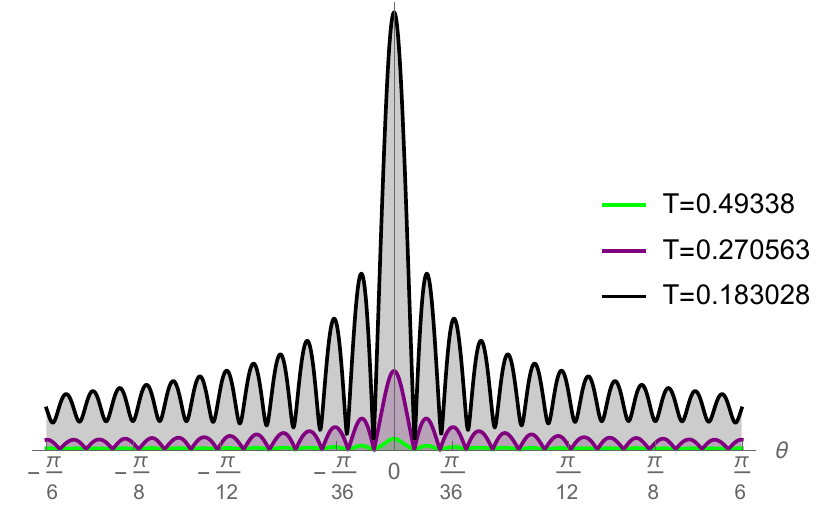}
	\caption{The amplitude of $|\bra \mathcal{O}\ket|$ for different  $T$  with $b=0.6$ and $\omega=80$.}
    \label{amplitude2}
	\end{minipage}
   
    \begin{minipage}{0.85\linewidth}
       \centering	\includegraphics[width=0.85\linewidth]{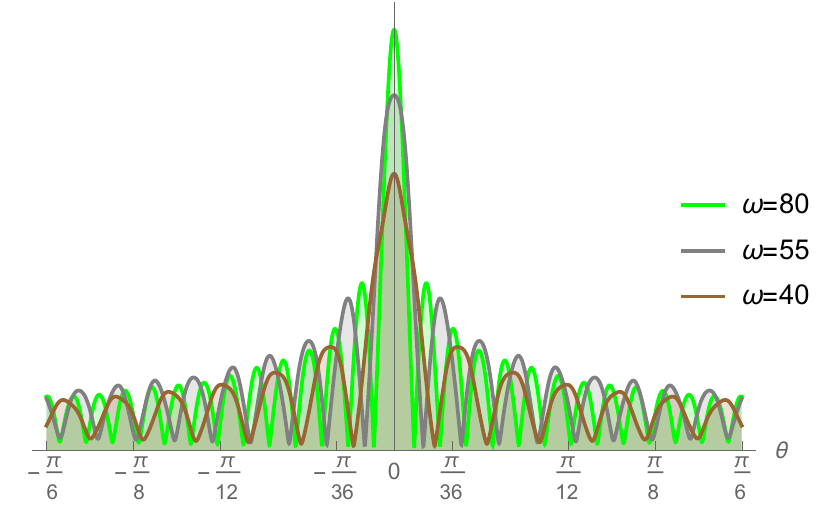}
	\caption{The amplitude of $|\bra \mathcal{O}\ket|$ for different $\omega$ with $b=0.6$ and $T=0.270563$.}
    \label{amplitude3}
    \end{minipage}
     \label{amplitude}
\end{figure}

The key task is to solve the radial equation Eq. (\ref{zpart}) with
the following boundary condition
\begin{equation}
	Z_l(0)=1
\end{equation}
at the AdS boundary condition and the regular boundary condition on the black hole event horizon. Here we employ the pseudo-spectral method~\cite{Liu2022} to obtain the corresponding numerical solution for $Z_l$ and extract $\mathcal{O}_l$. With the extracted $\mathcal{O}_l$, the total response can be obtained by Eq.~(\ref{resDecomp}). We plot a typical profile of the total response $|\bra\mathcal{O}\ket|$ in Fig.\ref{amplitude1} to Fig.\ref{amplitude3}. Closely followed by the method proposed in~\cite{Liu2022}, the interference pattern indeed arises from the diffraction of our scalar field off the black hole. For explicitly, in Fig.\ref{amplitude1}, the parameter of the monopole
$b$ changes, while the location of black hole event horizon $r_h=1$ and the frequency $\omega=80$ are fixed. Fig.~\ref{amplitude2} shows the amplitude of $|\mathcal{O}_l|$ for different temperature $T$ of the boundary system with the parameter of the monopole $b=0.6$ and the frequency $\omega=80$. Fig.~\ref{amplitude3} shows the amplitude of $|\bra \mathcal{O}\ket|$ for different  $\omega$ with $b=0.6$ and $T=0.270563$. These figures show that the absolute amplitude of total response function increases with the decrease of the monopole parameter $b$ and the temperature $T$. And more, the frequency $\omega$ of the Gaussian source will increase the absolute amplitude. In other words, the total response function depends closely on the Gaussian source and the spacetime geometry. Therefore, if this response function can be transformed as the observed images, it will be regarded as an useful tool to reflect the feature of the spacetime geometry. To achieve this goal, a special imaging system is required, which will be described specifically in the next section.

\section{Einstein ring in AdS black hole with a global monopole}
\label{sec4}
\begin{figure}[h]
	\centering
	\includegraphics[height=3.5in]{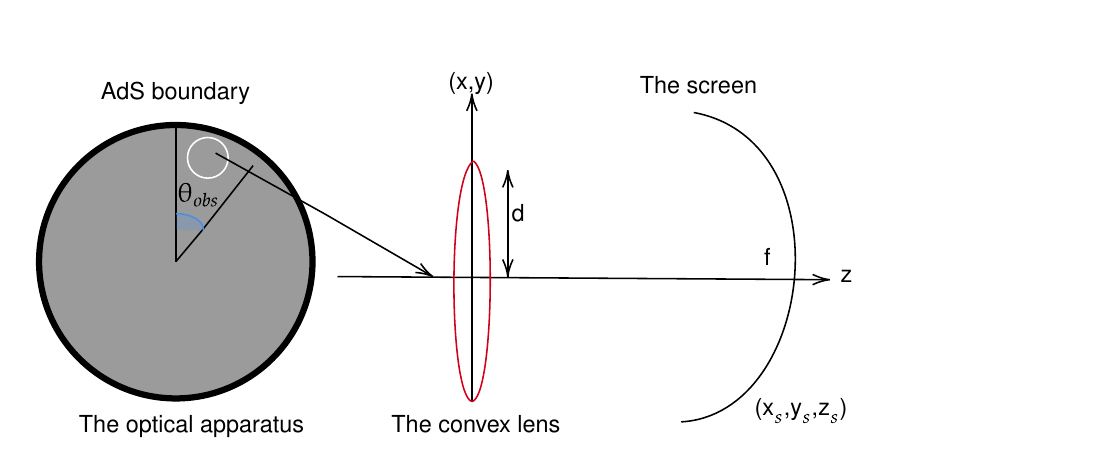}
	\caption{The observer and its telescope.}\label{telescope1}
\end{figure}

As stated above, we have observed the interference pattern resulting from the diffraction of the scalar wave by the black hole, which just suppose all amplitudes from different angles without differentiations. For observation, one should distinguish lights from different angles~\cite{Hashimoto:2019jmw,Hashimoto:2018okj}. By using a special imaging system, we are able to convert the above extracted response function $\bra\mathcal{O}\ket$ into the holographic image through an optical system named `telescope' shown in Fig.~\ref{telescope1}. Explicitly, the left represents the $(2+1)$-dimensional boundary CFT on the 2-sphere $S^2$ which is naturally dual to a black hole in the global $AdS_4$ spacetime. The middle is the convex lens with focal length f which is regarded as a "converter" between plane and spherical waves. Imagine that the incident wave from the left hand is irradiated at the lens, the wave will convert to the transmitted wave at the focus which will be received on the screen shown on the right side in Fig.~\ref{telescope1}. With this apparatus, we can explore whether the effect of the global monopole parameter in the spherically symmetric AdS solution is reflected in the holographic image characteristics or not, which in turn help us deeply understand the geometric structure of spacetime.

Explicitly we suppose an observer is surrounded by a small circle with the center located at $(\theta_{obs},0)$ on the unit sphere, which is shown in Fig.~\ref{telescope1}. By rotating the original spherical coordinate system $(\theta,\psi)$ to a new $(\theta^{\prime},\psi^{\prime})$ in such a way that
\begin{eqnarray}
\sin \theta^{\prime} \cos \psi^{\prime}= e^{i \theta_{obs}} (\sin \theta\cos\psi+i \cos\theta),
\end{eqnarray}
$(\theta^{\prime}=0,\psi^{\prime}=0)$ corresponds to the center of the observation region. For simplicity, we introduce a Cartesian coordinate system $(x_1,x_2,x_3)$ such that $(x_1,x_2)=(\theta^{\prime}\cos \psi^{\prime},\theta^{\prime}\sin\psi^{\prime})$ in the observation region. We set the convex lens on the $(x,y)$-plane. The focal length and radius of the lens will be denoted by $f$ and $d$. We adjust a spherical screen at $(x,y,z)=(x_s,y_s,z_s)$ with $x_s^2+y_s^2+z_s^2=f^2$.

\begin{figure}
    \centering
    \subfigure[$b=0.01$,$\theta_{obs}=0$]{
        \includegraphics[width=1.3in]{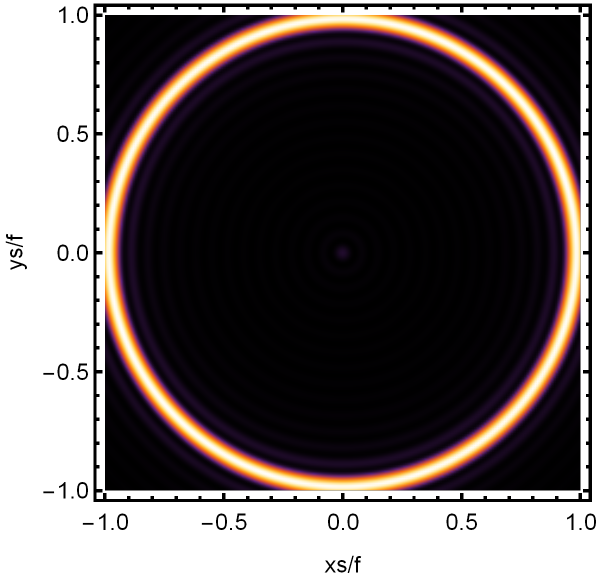}
    }
	\subfigure[$b=0.01$,$\theta_{obs}=\pi/6$]{
        \includegraphics[width=1.3in]{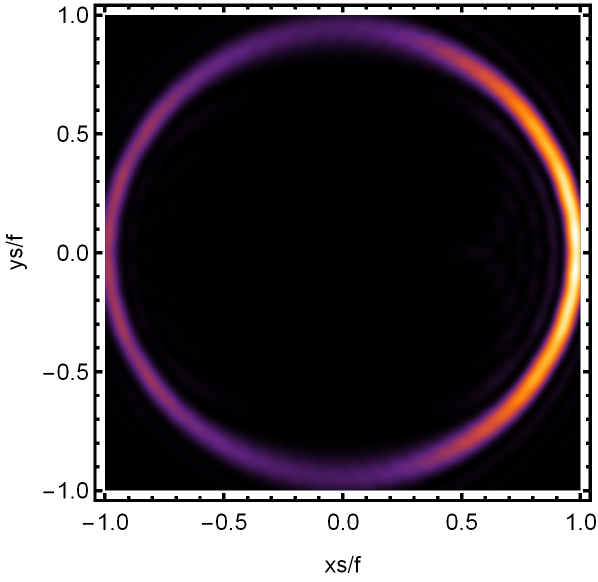}
    }
    \subfigure[$b=0.01$,$\theta_{obs}=\pi/3$]{
        \includegraphics[width=1.3in]{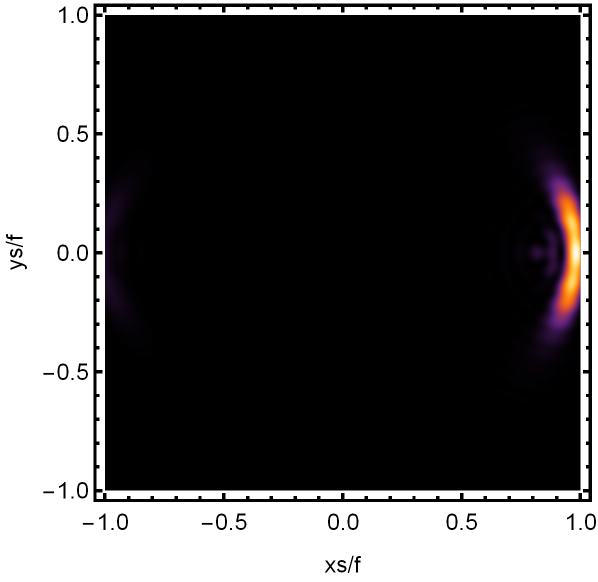}
    }
    \subfigure[$b=0.01$,$\theta_{obs}=\pi/2$]{
        \includegraphics[width=1.3in]{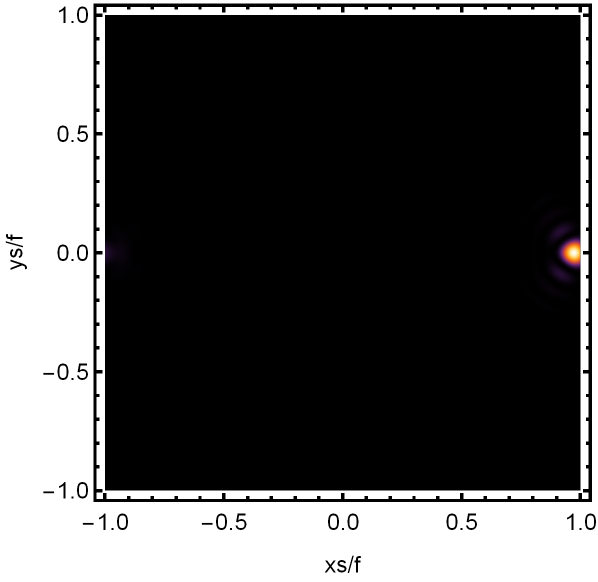}
    }
    \subfigure[$b=0.6$,$\theta_{obs}=0$]{
        \includegraphics[width=1.3in]{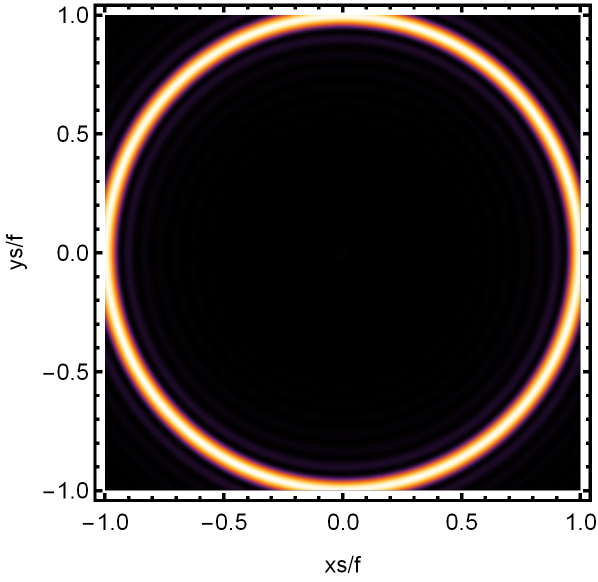}
    }
	\subfigure[$b=0.6$,$\theta_{obs}=\pi/6$]{
        \includegraphics[width=1.3in]{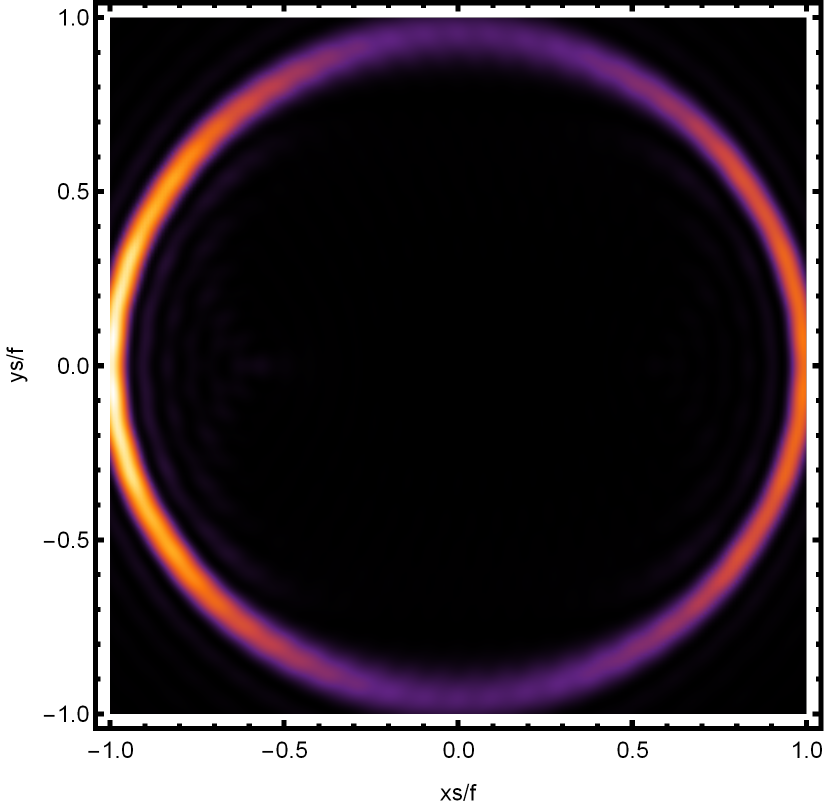}
    }
    \subfigure[$b=0.6$,$\theta_{obs}=\pi/3$]{
        \includegraphics[width=1.3in]{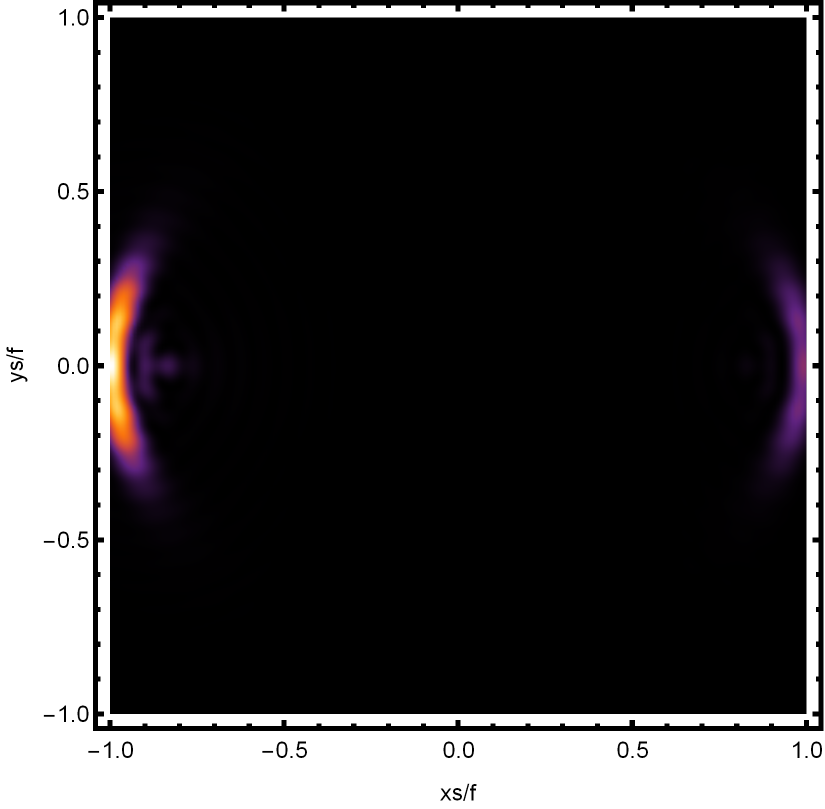}
    }
    \subfigure[$b=0.6$,$\theta_{obs}=\pi/2$]{
        \includegraphics[width=1.3in]{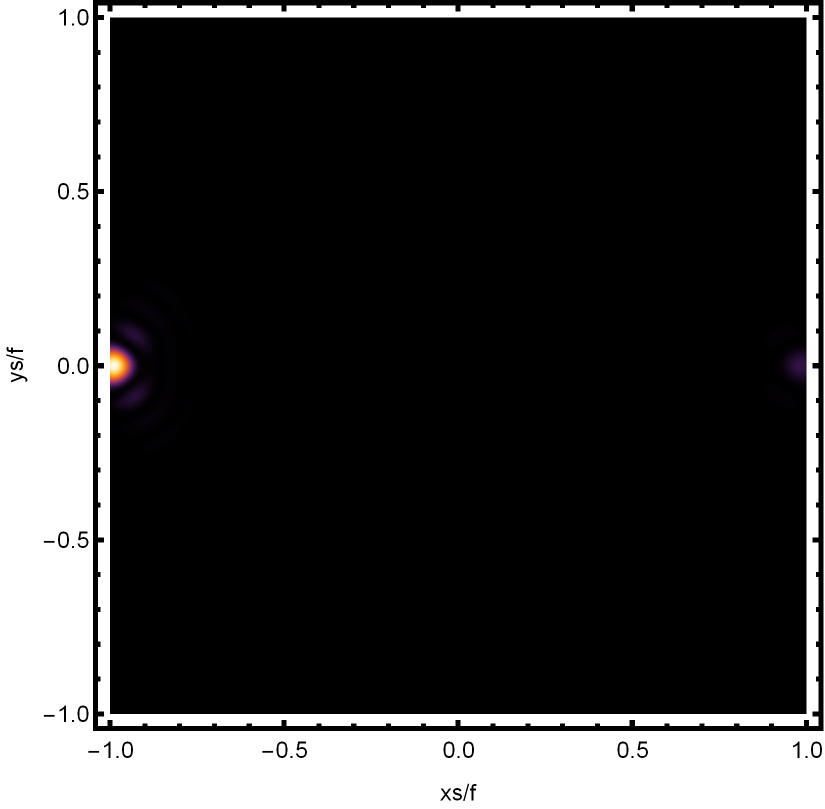}
    }
    \subfigure[$b=2$,$\theta_{obs}=0$]{
        \includegraphics[width=1.3in]{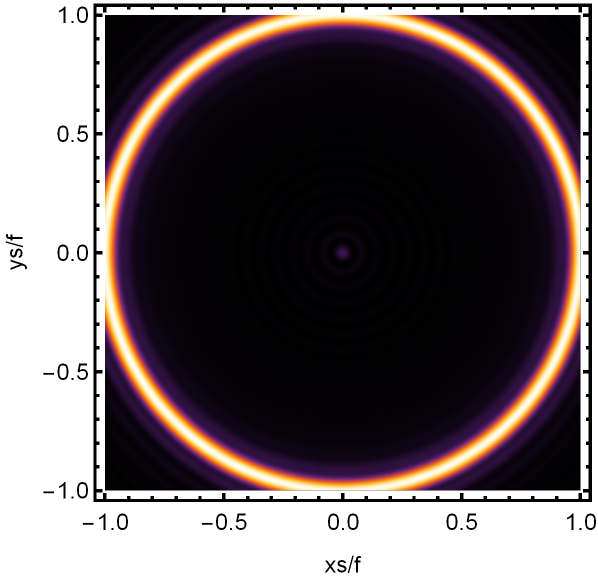}
    }
	\subfigure[$b=2$,$\theta_{obs}=\pi/6$]{
        \includegraphics[width=1.3in]{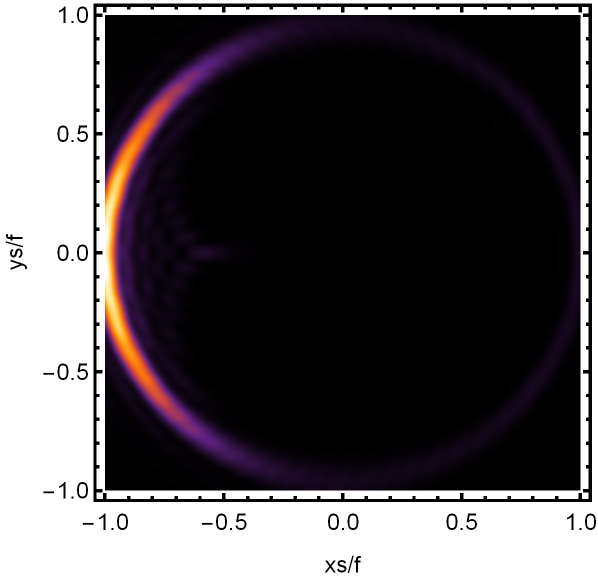}
    }
    \subfigure[$b=2$,$\theta_{obs}=\pi/3$]{
        \includegraphics[width=1.3in]{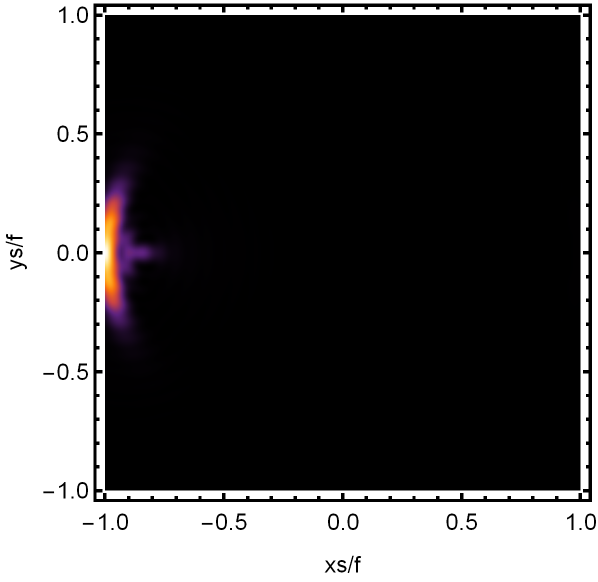}
    }
    \subfigure[$b=2$,$\theta_{obs}=\pi/2$]{
        \includegraphics[width=1.3in]{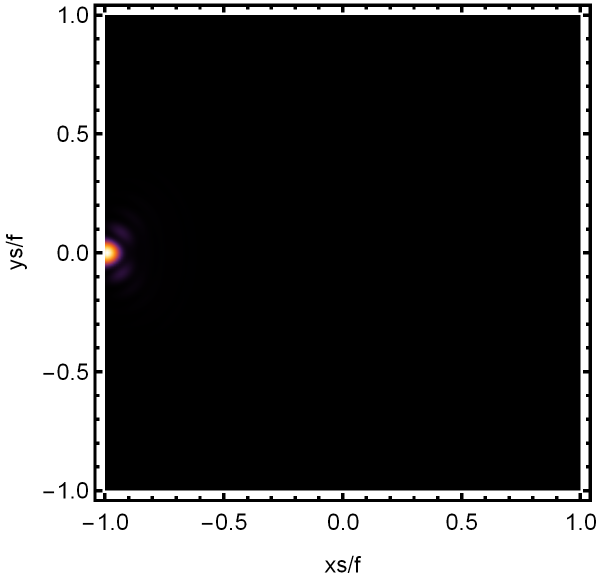}
    }
    \caption{The images of the lensed response observed at various observation angles for different $b$ with $T=0.270563$ and $\omega=80$.}\label{sharpimage1}
\end{figure}

Considering a wave $\Psi(\vec{x})$ with frequency $\omega$ getting through the convex lens, the transmitted wave $\Psi_T(\vec{x})$ should be
\begin{equation}
\Psi_T(\vec{x})=e^{-i \omega \frac{|\vec{x}|^2}{2f}}\Psi(\vec{x}).
\end{equation}
The role of the convex lens is to transmit a wave $\Psi(\vec{x})$ into a spherical wave. Then the wave function imaging on screen thus becomes
\begin{equation}\label{lensTranslation}
	\Psi_S(\vec x_S)=\int_{|\vec x|\smalleq d}\exd^2 x\Psi_T(\vec x)e^{i\omega D}
	\linear \int_{|\vec x|\smalleq d}\exd^2 x\Psi(\vec x) e^{-i\frac{\omega}{f}\vec x\cdot\vec x_S}= \int \exd^2 x\Psi(\vec x) w(\vec x) e^{-i\frac{\omega}{f}\vec x\cdot\vec x_S},
\end{equation}
where the integral is performed over the convex lens of radius $d$, $D$ is the propagating distance from the lens point $(x_1,x_2,0)$ to the screen point $(x_{S1},x_{S2},x_{S3})$, and $w(\vec x)$ is the window function, defined as
\begin{equation}
	w(\vec x):=\begin{cases}
		1,\quad 0\smalleq|\vec x|\smalleq d,\\
		0,\quad |\vec x|>d.
	\end{cases}
\end{equation}
According to Eq.(\ref{lensTranslation}), we clearly see that the observed wave on the screen connect with the incident wave by the Fourier transformation. We will capture the images of the dual black hole on the the screen with Eq.(\ref{lensTranslation}). When the observer located at different positions of AdS boundary, the holographic Einstein images have been presented in Fig.~\ref{sharpimage1} with the change of the global monopole parameter b and various observation angles for fixed $T=0.270563$ and $\omega=80$. Fig.~\ref{sharpimage2} shows the images of the lensed response observed at the observation angle $\theta_{obs}=0$  for different $T$ wit $b=0.6$ and $\omega=80$.

When the observer located at the position $\theta=0^{\circ}$, i.e., the observation location is the north pole of the AdS boundary, it can be seen that a series of axisymmetric concentric rings appear in the image, and one of them is particularly bright which is shown in the left column. From top to bottom, the monopole parameter b increases while the brightness of the ring is almost the same. Next we fix the observed position to $\theta=30^{\circ}$ (the second column from the left). When the monopole parameter b increases, the bright ring changed into a luminosity-deformed ring instead of a strict axisymmetric ring. And more, for smaller b, it is brighter at the right side. While b gets greater, the left side of ring gets brighter and the right side of ring is darker until the right side of the ring disappeared when $b=2$. Then we set the observation at $\theta=60^{\circ}$. There are just bright light arcs appeared which is consistent with~\cite{Hu:2020lkg}. And the bright light arcs exist on the right side when b is smaller and vice versa. When the observer is at $\theta=90^{\circ}$, all left is a bright spot shown on the right column of Fig.~\ref{sharpimage1}. We see a right bright spot when b is smaller and a left bright spot when b is bigger. Therefore, b have an effect on the position of the holographic Einstein image, which is used to distinguish different gravity theories.

\begin{figure}
	\centering
	\includegraphics[width=4.5in]{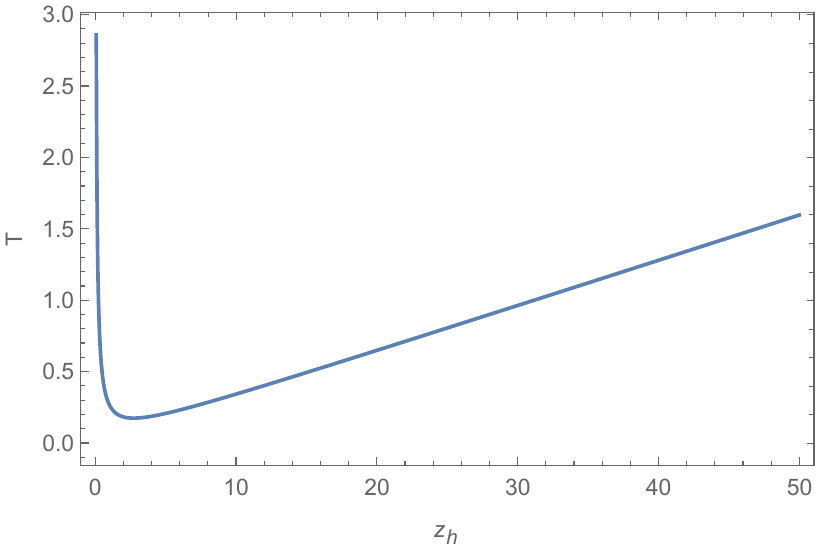}
	\caption{The relation between the temperature $T$ and inverse of  horizon $z_h$, with the minimum $(T=0.169955,z_h=2.12386)$.}
 \label{relation between temperature and horizon}
\end{figure}

For better understanding the physical meaning of the above holographic Einstein image, we study the impact of the horizon temperature on the images of the lensed response observed with the fixed observation angle, fixed monopole parameter and fixed frequency $\omega$ shown in Fig.~\ref{sharpimage2}. Before this, we plot the relation of the temperature and the horizon shown in Fig.~\ref{relation between temperature and horizon} with the lowest point $(T=0.169955,z_h=2.12386)$. It is clearly that with the increase of the event horizon, the temperature decreases at the beginning and then increases.

In Fig.~\ref{sharpimage2}, we plot the images of the lensed response observed with the fixed observation angle $\theta_{obs}=0$, fixed mpnopole parameter $b=0.6$ and $\omega=80$. From Fig.~\ref{relation between temperature and horizon}, it is easy to read out $T=2.39051$ whose corresponding horizon is $z_h=0.1$. $T=0.270563$ corresponds to $z_h=1$. And $T=0.342183$ corresponds to $z_h=10$. $T=1.59632$ means $z_h=50$. We study the image of the dual black hole when the horizon gradually increases. When the horizon $z_h=0.1$ is small, we see a very bright light spot in the center. When the horizon grows to $z_h=1$, we see a series of axisymmetric concentric rings appearing in the image, and one of them is particular bright and far away the center. When we further increase the horizon to $z_h=10$, the brightest ring becomes closer to the center and smaller. When $z_h=50$, we  see a bright spot in the center in addition to a bright and small ring closer to the center.

\begin{figure}
    \centering
    \subfigure[$T=2.39051$]{
        \includegraphics[width=1.4in]{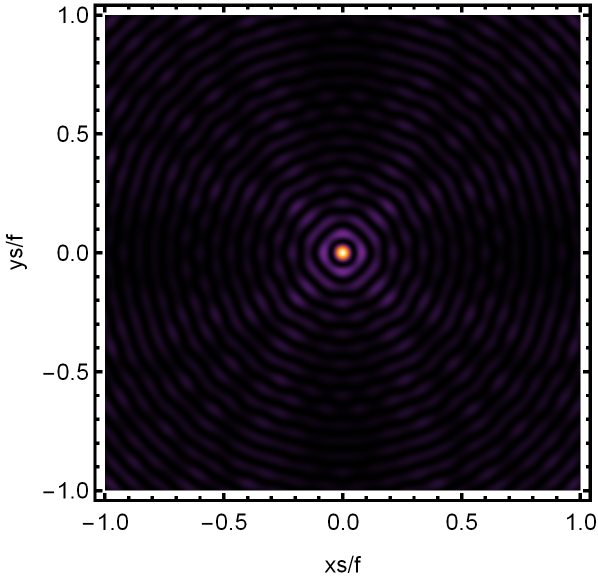}
    }
	\subfigure[$T=0.252044$]{
        \includegraphics[width=1.4in]{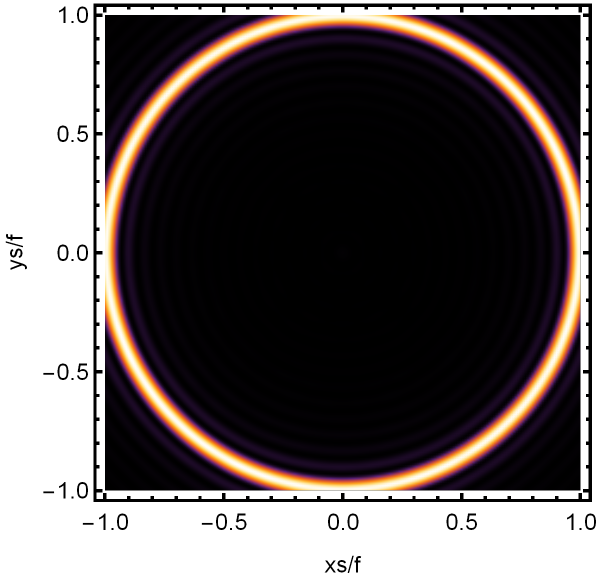}
    }
    \subfigure[$T=0.65168$]{
        \includegraphics[width=1.4in]{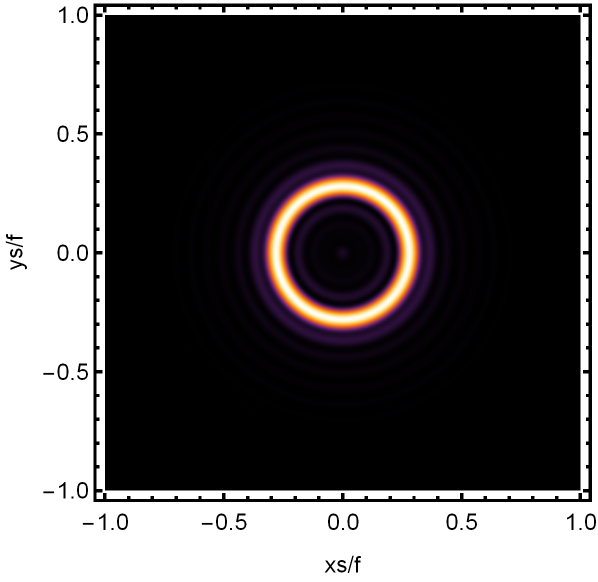}
    }
    \subfigure[$T=1.28238$]{
        \includegraphics[width=1.4in]{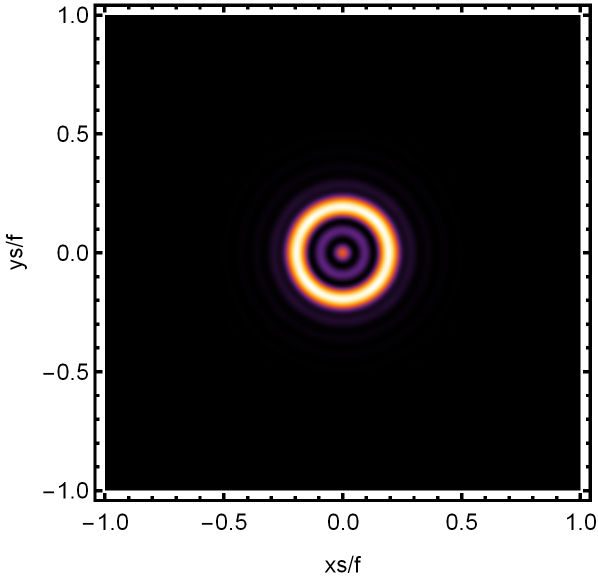}
    }
    \caption{The images of the lensed response observed at the observation angle $\theta_{obs}=0$  for different $T$ with $b=0.6$ and $\omega=80$.}\label{sharpimage2}
\end{figure}

\begin{figure}
    \centering
    \subfigure[$T=2.39051$]{
        \includegraphics[width=1.52in]{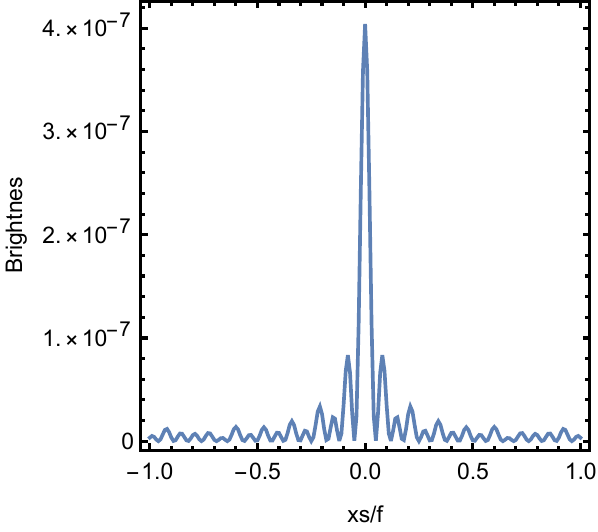}
    }
	\subfigure[$T=0.252044$]{
        \includegraphics[width=1.39in]{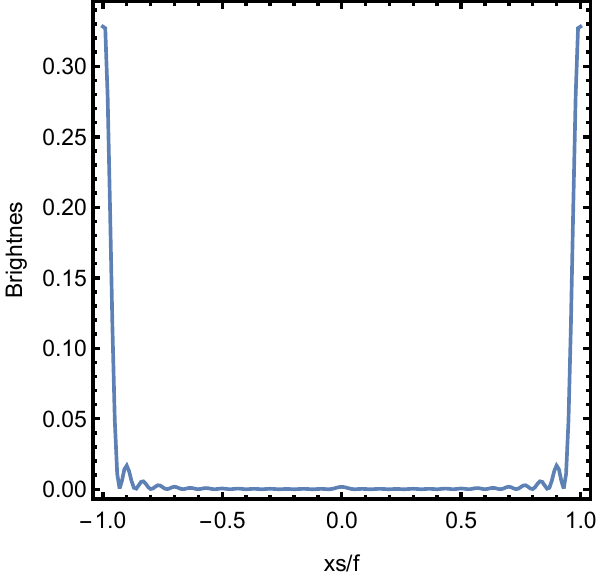}
    }
    \subfigure[$T=0.65168$]{
        \includegraphics[width=1.49in]{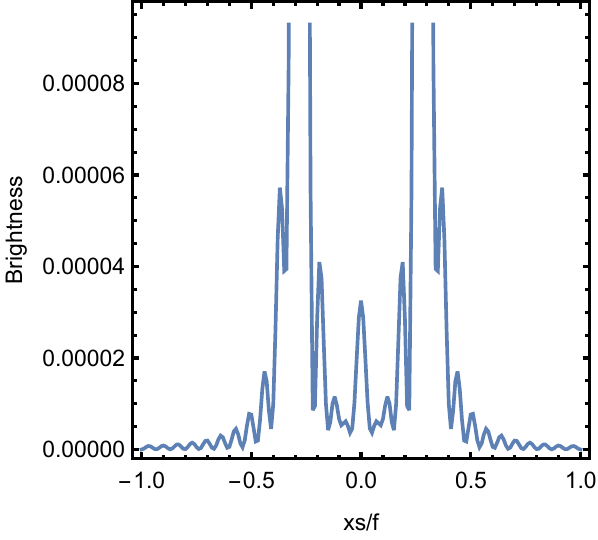}
    }
    \subfigure[$T=1.28238$]{
        \includegraphics[width=1.52in]{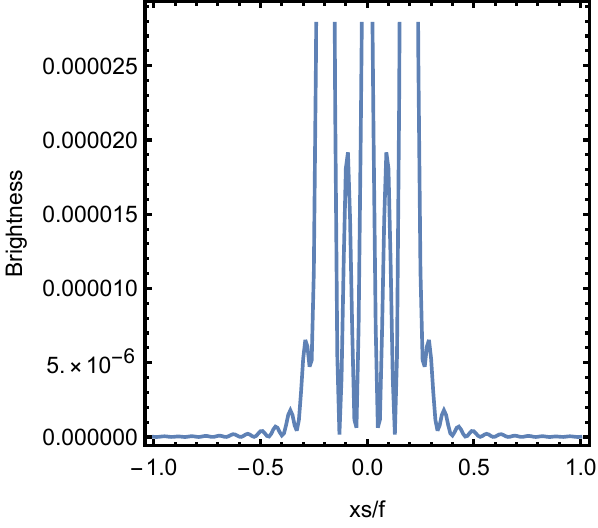}
    }
    \caption{The brightness of the lensed response on the screen for different $T$ with $\omega=80$ and $b=0.6$.}\label{sharpimage3}
\end{figure}

The above phenomena can be also seen  in Fig.~\ref{sharpimage3}. The subfigures in Fig.~\ref{sharpimage3} corresponds to those in Fig.~\ref{sharpimage2} one by one. From these subfigures in Fig.~\ref{sharpimage3}, we clearly see a peak in the center when $T=2.39051$ which corresponds to a light spot in Fig. 8a. And when $T=0.252044$, we see two peaks on both sides of the x-axis which corresponds to a bright ring in Fig. 8b. For $T=0.65168$ and $T=1.28238$, we see many peaks which correspond to a series of axisymmetric rings in Fig. 8c and 8d. From such figure,  when the temperature is lower, the brightest ring is located at focal point f. When the temperature T rises, the brightest ring gradually moves inward.

\begin{figure}
    \centering
    \subfigure[$b=0.51$]{
        \includegraphics[width=1.3in]{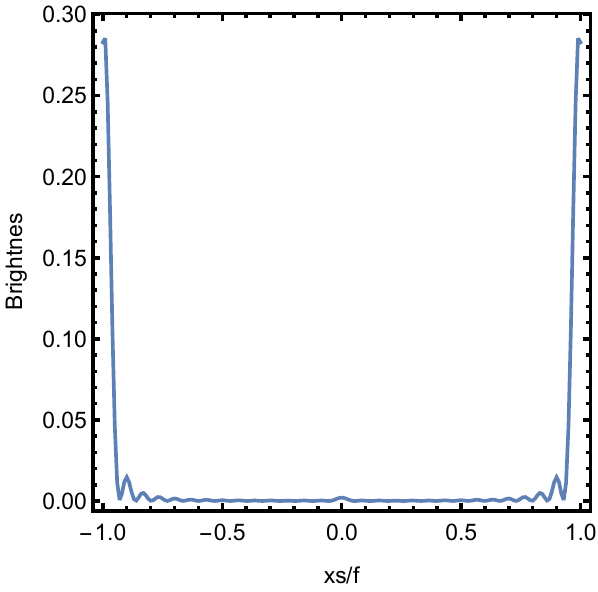}
    }
 \subfigure[$b=1.01$]{
        \includegraphics[width=1.3in]{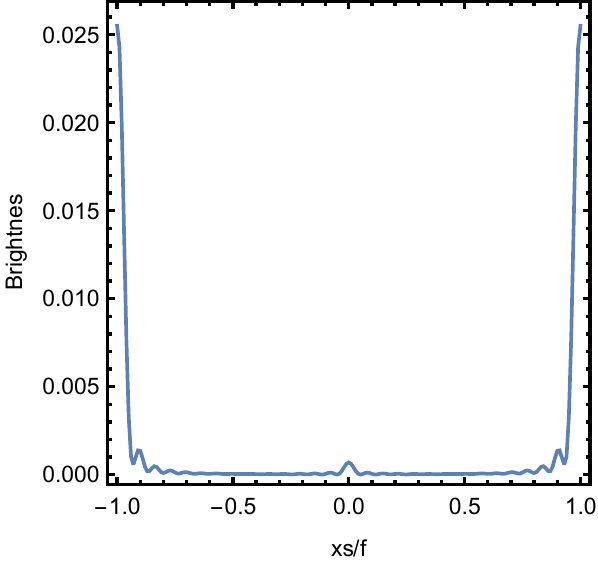}
    }
    \subfigure[$b=1.51$]{
        \includegraphics[width=1.3in]{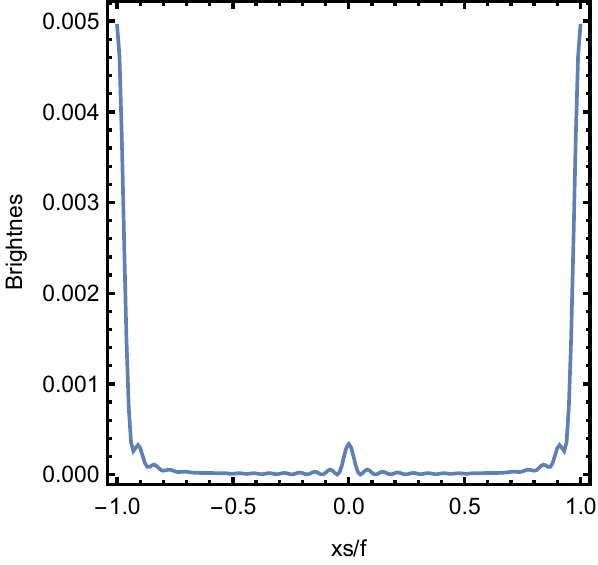}
    }
    \subfigure[$b=2.01$]{
        \includegraphics[width=1.3in]{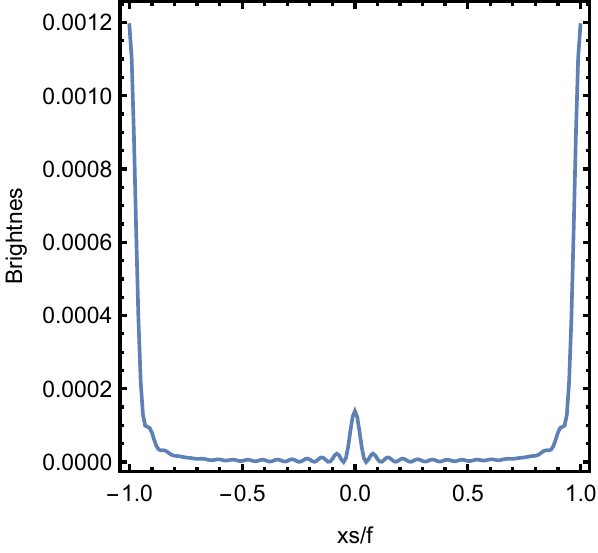}
    }
    \caption{ The brightness of the lensed response on the screen for different $b$ with $\omega=80$ and $z_h=1$.}
    \label{different b}
\end{figure}

And more, we further study the effect of the global monopole parameter on the image of the dual black hole. We draw curve plots in Fig.~\ref{different b}. The horizontal axis is the x-position and the vertical axis is the brightness of the lensed response on the screen for different b with the same $\omega=80$ and $z_h=1$. The parameter b is larger, the ring becomes dimmer. Therefore the holographic images of AdS black hole can not only characterize the geometric of black hole, but also closely related to the properties of lens and wave packet source.

\begin{figure}
    \centering
    \subfigure[$\omega=60$]{
        \includegraphics[width=1.3in]{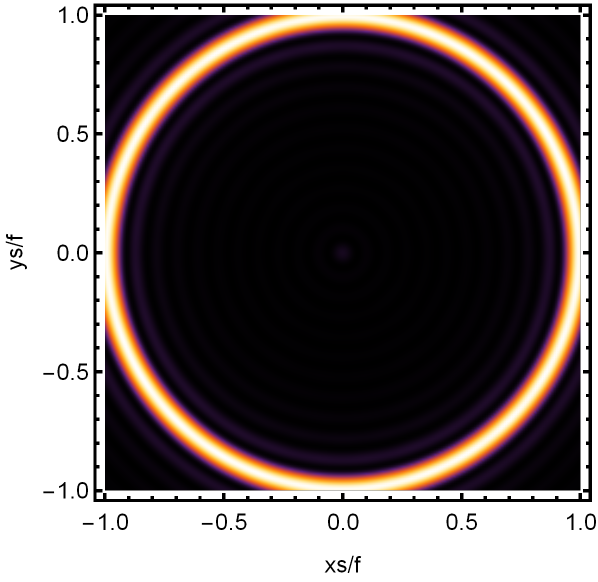}
    }
	\subfigure[$\omega=40$]{
        \includegraphics[width=1.3in]{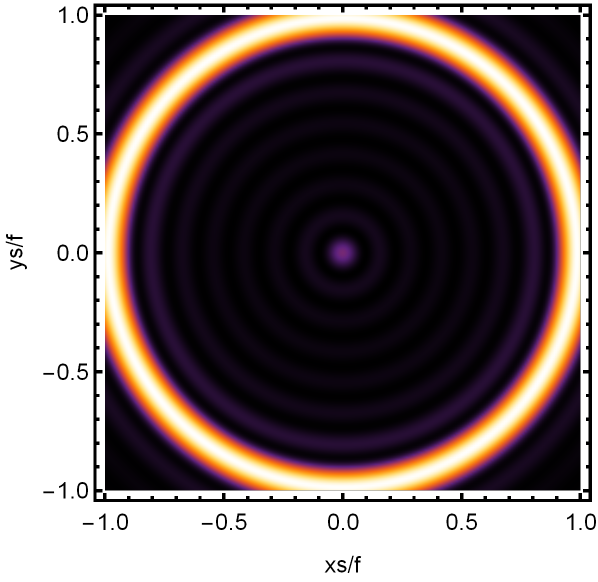}
    }
    \subfigure[$\omega=20$]{
        \includegraphics[width=1.3in]{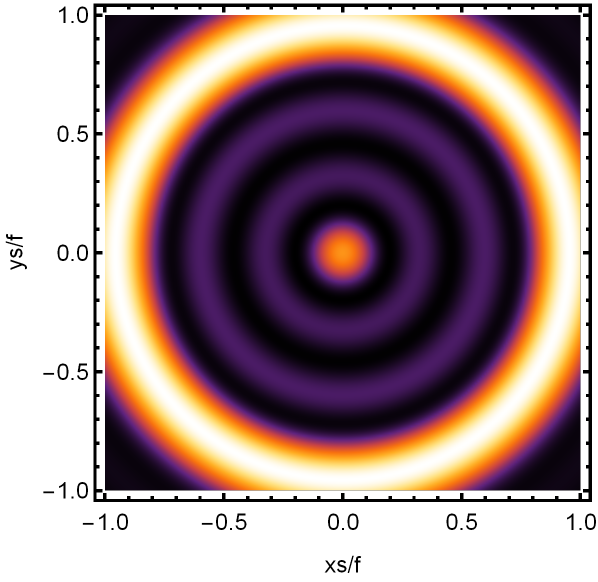}
    }
    \subfigure[$\omega=10$]{
        \includegraphics[width=1.3in]{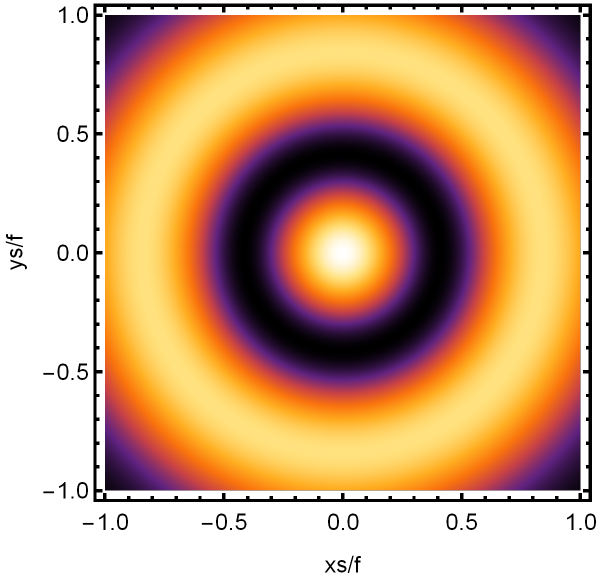}
    }
    \caption{The images of the lensed response observed at the observation angle $\theta_{obs}=0$  for different $\omega$ with $b=0.6$ and $T=0.27.563$.}\label{different_omega}
\end{figure}

In addition, we also consider the effect of wave source on the characteristics of the holographic Einstein image shown in Fig.~\ref{different_omega}. With $\sigma= 0.02$ for the source and $d=0.6$ for the convex lens, the higher the frequency becomes, the sharper the resulting ring becomes. This is reasonable because the image can be well captured by the geometric optics approximation in the high frequency limit. 

For the brightest ring in the image, it corresponds to the position of the photon sphere of the black hole. Next we will verify this bright ring in the image from the perspective of optical geometry. In a spacetime with metric in Eq.(\ref{metric1}), we can express the ingoing angle of photons from boundary with the conserved energy $\omega$ and the angular momentum $L$. Without loss of generality, we choose the coordinate system in order to let the photon orbit lying on the equatorial plane $\theta=\pi/2$.
The 4-vector $u^a=(d/d\lambda)^a$ in satisfies
\begin{equation}
-F(r)\left(\diff{t}{\lambda}\right)^2+\frac{1}{F(r)}\left(\diff{r}{\lambda}\right)^2+r^2\sin^2\theta\left(\diff{\phi}{\lambda}\right)^2=0,
\end{equation}
or equivalently,
\begin{equation}
	\dot r^2=\omega^2-L^2\mathcal{R},
\end{equation}
where $\mathcal{R}=F(r)/r^2$, $\omega =F(r)\dot{t}$,  $L=r^2 \dot{\phi}$, and  $\dot r\notice \pd r/\pd\lambda$, $\dot t\notice \pd t/\pd\lambda$, $\dot \phi\notice \pd \phi/\pd\lambda$.

Its ingoing angle $\theta_\text{in}$ with normal vector of boundary $n^a=\pd/\pd{r}^a$ should be
\begin{equation}
	\cos\theta_\text{in}=\frac{g_{ij}u^in^j}{|u||n|}\when_{r=\infty}=\sqrt{\frac{\dot r^2/F}{\dot r^2/F+L^2/r^2}}\when_{r=\infty},
\end{equation}
which means that
\begin{equation}
	\sin^2\theta_\text{in}=1-\cos^2\theta_\text{in}=\frac{L^2 \mathcal{R}}{\dot r^2+L^2\mathcal{R}} \when_{r=\infty}=\frac{L^2}{\omega^2}.
\end{equation}

\begin{figure}[h]
\includegraphics[trim=0.4cm 0.3cm 5.4cm 2.5cm, clip=true, scale=0.7]{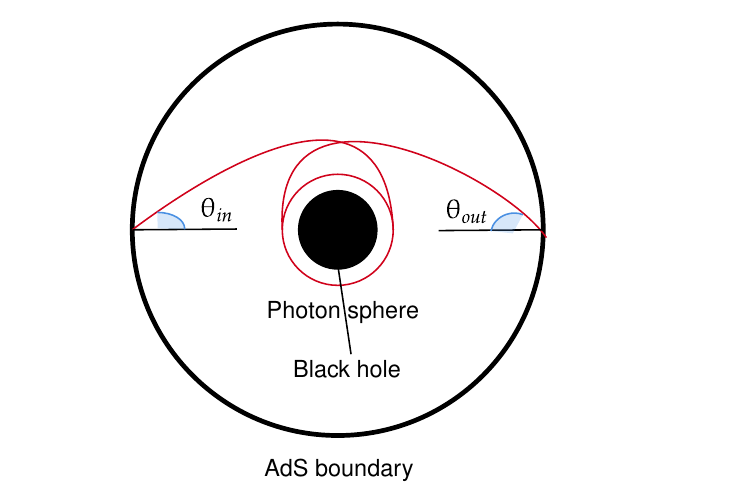}
\caption{The dominant contribution to the final response function comes from the trajectory as close to the circular orbit as possible.}\label{trajectory}
\end{figure}

So the ingoing angle $\theta_\text{in}$ of photon orbit from boundary satisfies that
\begin{equation}
	\sin\theta_{in}=\frac{L}{\omega},
\end{equation}
which is shown in Fig.~\ref{trajectory}. Especially, when  the light is located at the photon sphere, this relation is still valid. We label the angular momentum in the case as $L_s$, which is determined by the following conditions
\begin{eqnarray}
\mathcal{R}=0,   \ \ \   \frac{d\mathcal{R}}{dr}=0.
\end{eqnarray}

In the geometrical optics, the angle $\theta_{in}$ gives the angular distance of the image of the incident ray from the zenith if an observer on the AdS boundary looks up into the AdS bulk. If two end points of the geodesic and the center of the black hole are in alignment, the observer see a ring with a radius corresponding to the incident angle $\theta_{in}$ because of axisymmetry~\cite{Hashimoto:2018okj}.

\begin{figure}[h]
\includegraphics[trim=0.4cm 2.0cm 0.0cm 0.0cm, clip=true, scale=0.8]{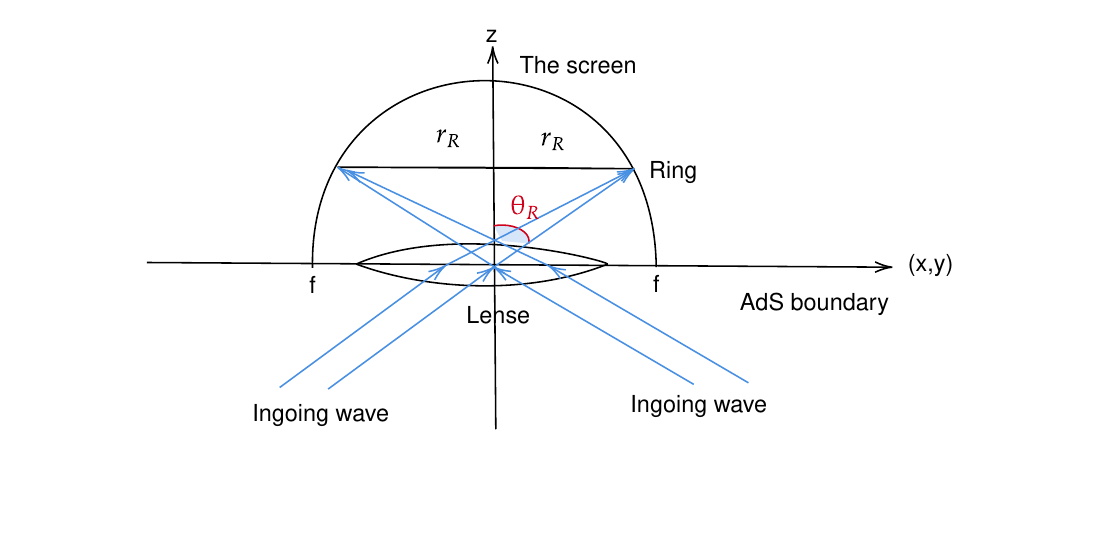}
\caption{The relation between the ring radius and ring angle.}\label{ingoing1}
\end{figure}

\begin{figure}
	\centering
	\begin{minipage}[t]{0.48\textwidth}
	\centering
	\includegraphics[height=2in]{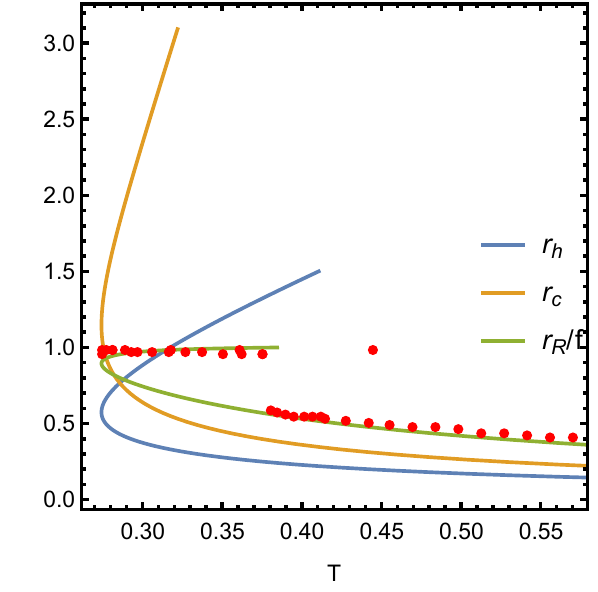}
	\caption{The circular orbit $r_c$, event horizon $r_h$, and ring radius $r_R$ in the unit of $f$ as functions of temperature $T$ with  $\omega=80$ and $b=0.01$.}\label{radius11}
	\end{minipage}
       \begin{minipage}[t]{0.48\textwidth}
	\centering
	\includegraphics[height=2in]{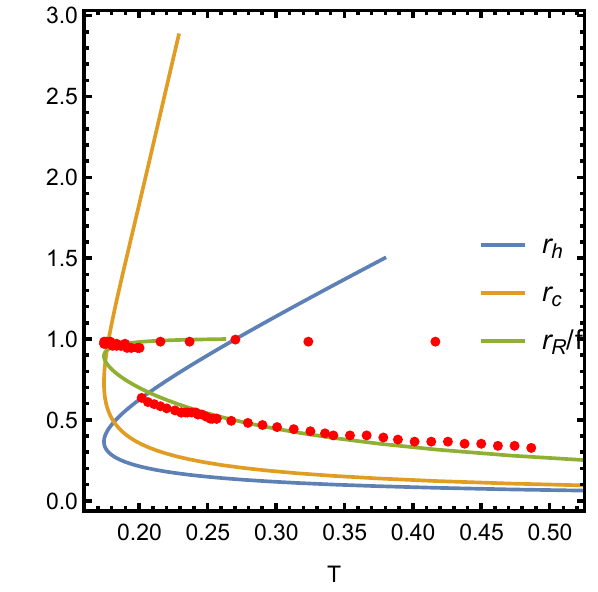}
	\caption{The circular orbit $r_c$, event horizon $r_h$, and ring radius $r_R$ in the unit of $f$ as functions of temperature $T$ with  $\omega=80$ and $b=0.6$.}\label{radius12}
	\end{minipage}
\end{figure}

In addition, with Fig.~\ref{ingoing1}, we can
obtain the angle of the Einstein ring, that is  
\begin{equation}
	\sin\theta_{R}=\frac{r_R}{f}.
\end{equation}
According to \cite{Hashimoto:2018okj}, we know that for a sufficiently large $l$, $\sin\theta_{R}=\sin\theta_{in}$, thus we have the relation 
\begin{equation}
\frac{r_{R}}{f}=\frac{L_s}{\omega}. 
\end{equation}
This relation also can be confirmed numerically.
In particular, the value of $\frac{r_R}{f}$ of the Einstein ring formed are shown in Fig.~\ref{radius11} and Fig.~\ref{radius12}, where both radii of the black hole horizon $r_h$ and the circular orbit $r_R$ as a functions of temperature are exhibited. From Fig.~\ref{radius11}, the appreciable increase in the Einstein ring radius occurs only at low temperatures. After this increase, the Einstein ring radius starts to flatten out. And more, Fig.~\ref{radius12} tells us that the Einstein ring radius keeps almost unchanged while the inflection point shifts to the left when we increase the monopole parameter b from $b=0.01$ to $b=0.6$. As expected, the Einstein ring radius obtained by our wave optics fits well with that by geometric optics.

\section{Conclusions}
\label{sec5}
Gravitational lensing is a fundamental phenomena of strong gravity. The observers will see the Einstein ring when a light source is behind a gravitational body. In recent years, the EHT has captured the first image of the supermassive black hole in M87. Einstein ring contains a lot of information and the study of Einstein ring enables us not only to understand the geometric structure of spacetime, but also helps us explore the gravity models deeply. Therefore, in the framework of AdS/CFT correspondence, we have studied the holographic Einstein images of a spherically symmetric AdS black hole with a global monopole in the bulk. We consider a $(2+1)$ dimensional boundary conformal field theory on a 2-sphere $S^2$ at a finite temperature, and study a one-point function of a scalar temperature $\mathcal{O}$, under a time-dependent localized Gaussian source $J_{\mathcal{O}}$ with the frequency $\omega$. We derive the local response function $e^{-i \omega t}\mathcal{O}(\bar{x})$. With this given response function, our results show that the radius of the Einstein ring is dependent on the change of temperature or say the positions of horizon. And more, When the positions of observer varies, the Einstein ring will change into a luminosity-deformed ring, light arcs, or a central light spot or a combination of them. We also verify the brightest ring corresponds to the position of the photon sphere from the perspective of optical geometry, which further implies the position of holographic Einstein ring is full consistent with that of geometrical optics.

Different gravitational theories have different effects on the Einstein ring. Here we carefully study the effect of the global monopole parameter on the Einstein ring. We find that the monolople parameters can change the brightness of the ring but has a slight effect on the positions of the ring.  
These results maybe imply that the holographic images can be used as an effective tool to distinguish different types of black holes for fixed wave source and optical system. Obviously, it is also very interesting to further study the holographic images in other modified gravity theories. And it is also interesting to extend this method to the case of other different fields.

\section*{Acknowledgements}{Li-Fang Li would like to thank Haiqing Zhang for his help. This work is supported  by the National
Natural Science Foundation of China (Grants No. 11675140, No. 11705005, and No. 11875095), and  Innovation and Development Joint  Foundation of Chongqing Natural Science  Foundation (Grant No.CSTB2022NS
CQ-LZX0021).}

\end{document}